\documentclass[11pt]{article}

\usepackage[margin=1in]{geometry}
\usepackage{amsmath,amssymb,amsthm,bm,mathtools}
\usepackage{graphicx}
\usepackage{booktabs}
\usepackage{array}
\usepackage{enumitem}
\usepackage{microtype}
\usepackage{textcomp}
\usepackage[hidelinks]{hyperref}
\hypersetup{
  pdftitle={Initial-State Precision as a Predictive Resource: From Tori to Strange Nonchaotic Attractors and Chaos},
  pdfauthor={Song-Ju Kim}
}

\newtheorem{proposition}{Proposition}

\newcommand{\R}{\mathbb{R}}
\newcommand{\T}{\mathbb{T}}

\newcommand{\eps}{\varepsilon}

\title{\textbf{Initial-State Precision as a Predictive Resource: From Tori to Strange Nonchaotic Attractors and Chaos}}

\author{
Song-Ju Kim\\
\small SOBIN Institute LLC, Kawanishi, Hyogo, Japan\\
\small \texttt{kim@sobin.org}
}

\date{\today}

\begin{document}
\maketitle

\begin{abstract}
How much initial-state precision is required to predict a nonlinear system to a prescribed
accuracy over a finite horizon? We formulate this inverse prediction problem through a
one-shot resource $B_N$, defined as the number of binary refinement bits required in the
initial state by a specified local sensing architecture. Within a common scale-separable
error-growth regime, changing a fixed tolerance changes $B_N$ only by an $N$-independent
offset. An exactly solvable circle/Chebyshev map provides the chaotic calibration
$B_N=N\log_2 m+O(1)$.

We then apply the framework to the quasiperiodically forced logistic map under phase-only initial uncertainty. Over the resolved horizons, representative smooth-torus, strange-nonchaotic-attractor (SNA), and chaotic regimes exhibit a three-level precision-resource hierarchy: bounded, logarithmic-like, and linear-like growth,
respectively. For the SNA, algebraic phase sensitivity $G_N^{\mathrm{op}}\sim N^\mu$ implies an admissible initial uncertainty of order $N^{-\mu}$, and therefore a precision cost proportional to $\log N$. Phase-grid refinement shows that the sampled worst-case maxima are numerically stable with increasing phase resolution. Near the torus--SNA fractalization point, derivative-based phase sensitivity can already grow while finite perturbations remain below a fixed operational tolerance for more than $10^6$ iterations, revealing exceptionally long activation horizons. Thus an SNA can occupy an intermediate finite-precision resource class between a smooth torus and ordinary chaos, while its operational manifestation can occur on a much longer timescale than its derivative-based sensitivity.
\end{abstract}

\section{Introduction}

Prediction is often characterized by forward error growth: nearby states separate,
finite-size perturbations amplify, and Lyapunov exponents or related quantities
summarize that instability.  A physical measurement problem is naturally posed in the
opposite direction.  Given a target, an allowed error, and a finite prediction horizon,
how accurately must the initial condition be specified?

We address that inverse problem with a deliberately restricted resource architecture.
The resource is the local physical resolution of the initial state or of a declared
initial coordinate.  It is not an unrestricted source code: an encoder that already
knows the exact state and may transmit the future target directly would bypass the
dynamical prediction problem.  The question is instead how the required \emph{initial
measurement precision} scales with the horizon.

This viewpoint is related to finite-size Lyapunov analysis and
$\varepsilon$-dependent entropy \cite{Cencini2000,Boffetta1998}, predictive
rate-distortion theory \cite{MarzenCrutchfield2016}, metric mean dimension
\cite{LindenstraussTsukamoto2018}, and finite-rate state estimation
\cite{Kawan2021}, but the resource considered here is different: all of it is spent at
the initial time.  A complementary formulation optimizes the \emph{shape} of a fixed
uncertainty budget among predictive directions \cite{Kim2026RIPG}; here we ask how the
\emph{amount} of required resolution grows with prediction horizon.

Strange nonchaotic attractors (SNAs) provide a particularly useful test.
They are geometrically strange while the largest nontrivial Lyapunov exponent can
remain nonpositive \cite{Grebogi1984,PrasadNegiRamaswamy2001}.  At the same time,
their response to the phase of quasiperiodic forcing can become unbounded, as
quantified by the phase-sensitivity construction of Pikovsky and Feudel
\cite{PikovskyFeudel1995}.  Thus a smooth torus, an SNA, and ordinary chaos provide
three qualitatively different forms of sensitivity against which an initial-precision
resource can be calibrated.

The paper is organized around that comparison.  We first define a horizon-wide
one-shot precision resource and record two elementary consequences needed for its
interpretation.  An exactly solvable expanding circle/Chebyshev system then provides
a closed-form chaotic calibration.  The main numerical result is obtained for the
quasiperiodically forced logistic map: representative torus, SNA, and chaotic regimes
show bounded, logarithmic-like, and linear-like precision growth, respectively.
We then explain why algebraic phase sensitivity produces logarithmic precision demand,
test the numerical worst-case approximation with phase grids up to thousands of
samples, and finally study the very long activation horizons near the torus--SNA
onset.  Additional accuracy crossovers and broader parameter scans are reported in the Appendices.

\section{Initial precision as an operational resource}

Let $(X,d_X)$ be a metric state space, $F:X\to X$ a deterministic one-step map,
and $g:X\to\R$ the scalar prediction target.  Starting from $x_0\in X$,
\begin{equation}
x_{n+1}=F(x_n),\qquad
y_n=g(F^n x_0).
\end{equation}
For a horizon $N$, define the target trajectory
\begin{equation}
\bm Y_N(x)=\bigl(g(x),g(Fx),\ldots,g(F^N x)\bigr)\in\R^{N+1}
\end{equation}
with
\begin{equation}
\|\bm y-\bm y'\|_\infty=\max_{0\le n\le N}|y_n-y_n'|.
\end{equation}
A sensing architecture is specified by a nested family of local uncertainty cells
$\mathcal U_\delta(x)\subset X$, with $x\in\mathcal U_\delta(x)$ and larger
$\delta$ representing coarser initial information.  A metric ball
$\mathcal U_\delta(x)=\{x':d_X(x,x')\le\delta\}$ is the standard isotropic
example, but the formulation also allows a declared coordinate-selective sensor, such
as phase-only uncertainty in the forced system below.  The resource is therefore always
conditional on the stated sensing architecture.

Three associated error quantities are useful.  The worst-case trajectory diameter is
\begin{equation}
D_N(\delta)
=
\sup_{x\in X}
\operatorname{diam}_{\infty}\bm Y_N(\mathcal U_\delta(x)).
\label{eq:D}
\end{equation}
The minimax prediction radius is
\begin{equation}
E_N(\delta)
=
\sup_{x\in X}
\inf_{\widehat{\bm y}\in\R^{N+1}}
\sup_{x'\in\mathcal U_\delta(x)}
\|\bm Y_N(x')-\widehat{\bm y}\|_\infty,
\label{eq:E}
\end{equation}
where $\widehat{\bm y}$ is the trajectory estimate chosen for the uncertainty cell.
If instead the trajectory from the central state is used as predictor, the
nominal-trajectory error is
\begin{equation}
R_N(\delta)
=
\sup_{x\in X}
\sup_{x'\in\mathcal U_\delta(x)}
\|\bm Y_N(x')-\bm Y_N(x)\|_\infty.
\label{eq:R}
\end{equation}
They obey
\begin{equation}
\frac12D_N(\delta)\le E_N(\delta)\le R_N(\delta)\le D_N(\delta)\le2R_N(\delta).
\label{eq:DER}
\end{equation}
For scalar trajectory sets under the sup norm, $E_N=D_N/2$.

Fix a reference half-width $\delta_{\rm ref}>0$.  For a minimax tolerance
$\eps>0$, define
\begin{equation}
\delta_N^*(\eps)
=
\sup\{\delta\le\delta_{\rm ref}:E_N(\delta)\le\eps\}
\end{equation}
and the one-shot initial precision resource
\begin{equation}
\boxed{
B_N(\eps)
=
\left[
\log_2\frac{\delta_{\rm ref}}{\delta_N^*(\eps)}
\right]_+.
}
\label{eq:B}
\end{equation}
For a one-dimensional uniform sensor, $B_N$ is the additional number of binary
resolution bits required relative to $\delta_{\rm ref}$, up to integer rounding.
The full horizon dependence is retained because a linear rate
$\limsup B_N/N$ cannot distinguish bounded from logarithmic or other sublinear
growth.

\subsection{Scale-separable resource law}

The following proposition provides an organizing calibration.

\begin{proposition}[Scale-separable resource law]
\label{prop:scale}
Suppose that, for all sufficiently large $N$ and all relevant sufficiently small
$\delta$, there exist $p>0$, constants $0<c_-\le c_+<\infty$, and a positive
amplification function $G(N)$ such that
\begin{equation}
c_-G(N)\delta^p
\le E_N(\delta)
\le c_+G(N)\delta^p.
\label{eq:separable}
\end{equation}
Then every fixed accuracy whose optimal resolution eventually remains in this
regime satisfies
\begin{equation}
B_N(\eps)
=
\frac1p\log_2G(N)
+
\frac1p\log_2\frac1\eps
+
O(1).
\label{eq:scaleB}
\end{equation}
Hence two fixed accuracies in the same regime differ by only $O(1)$ bits.
\end{proposition}

The proof is immediate by inverting the two inequalities in
Eq.~\eqref{eq:separable}.  The point is operational rather than algebraic:
accuracy cannot change the horizon-scaling class while the same uniform
scale-separable law remains active.  A class change therefore signals departure
from that common regime, not merely a different numerical value of the tolerance.

The hierarchy follows directly:
\begin{equation}
G(N)=O(1)\Rightarrow B_N=O(1),\qquad
G(N)\sim N^\mu\Rightarrow B_N=\mu p^{-1}\log_2N+O(1),
\end{equation}
while $G(N)\sim e^{\lambda N}$ gives
\begin{equation}
B_N=\frac{\lambda}{p\ln2}N+O(1).
\label{eq:hierarchy}
\end{equation}

Define
\begin{equation}
E_\infty(\delta)=\sup_{N\ge0}E_N(\delta),\qquad
\eps_c=\inf_{0<\delta\le\delta_{\rm ref}}E_\infty(\delta).
\end{equation}
Then $\eps>\eps_c$ implies $B_N(\eps)=O(1)$, whereas
$\eps<\eps_c$ implies $B_N(\eps)\to\infty$.  This bounded/unbounded separation is
closely related to familiar sensitivity and equicontinuity scales; what matters
here is the growth law of $B_N$ below the scale.

\section{Exactly solvable circle/Chebyshev calibration}

Consider the expanding circle map
\begin{equation}
\theta_{n+1}=m\theta_n\pmod1,\qquad
\theta_n\in\T=\R/\mathbb Z,\qquad m\ge2,
\label{eq:circle}
\end{equation}
with the circle metric
\begin{equation}
d_{\T}(\theta,\theta')=\min_{k\in\mathbb Z}|\theta-\theta'-k|
\end{equation}
and observable
\begin{equation}
g(\theta)=a\cos(2\pi\theta),\qquad a>0.
\end{equation}
For this calibration the declared uncertainty cell is the phase ball
$\mathcal U_\delta(\theta)=\{\theta':d_{\T}(\theta,\theta')\le\delta\}$.
Thus two admissible phases may differ by as much as $2\delta$.

If $m^N\delta\le1/4$, the exact horizon-wide output diameter is
\begin{equation}
D_N(\delta)=2a\sin(2\pi m^N\delta).
\label{eq:circleD}
\end{equation}
Once $m^N\delta\ge1/4$, the uncertainty interval contains a half-turn separation
at or before horizon $N$, and the cosine reaches its full diameter $2a$.
For the minimax architecture, therefore, every fixed fine tolerance $0<\eps<a$
has
\begin{equation}
\delta_N^*(\eps)
=
\min\!\left\{
\delta_{\rm ref},
\frac{1}{2\pi m^N}
\arcsin\!\left(\frac{\eps}{a}\right)
\right\}
\end{equation}
and, for fixed $0<\delta_{\rm ref}\le1/4$,
\begin{equation}
\boxed{
B_N(\eps)
=
\left[
N\log_2m+
\log_2\!\left(
\frac{2\pi\delta_{\rm ref}}
{\arcsin(\eps/a)}
\right)
\right]_+.
}
\label{eq:chebB}
\end{equation}
Thus all fixed fine accuracies eventually have the same linear resource class and
differ by a constant number of bits.

The familiar Chebyshev form follows by defining
\begin{equation}
x_n=\cos(2\pi\theta_n).
\end{equation}
Using $T_k(\cos u)=\cos(ku)$,
\begin{equation}
x_{n+1}=T_m(x_n),\qquad
x_n=T_{m^n}(x_0)=\cos(2\pi m^n\theta_0).
\label{eq:cheb}
\end{equation}
The orbit is therefore available in closed form.  Here the Chebyshev representation serves as an exactly solvable calibration of the initial-precision resource in ordinary chaos: exponential instability gives $B_N=O(N)$, while changing a fixed fine accuracy changes only the additive offset.

\section{Quasiperiodically forced logistic map}

We study
\begin{align}
x_{n+1}
&=
\alpha[1+\epsilon_{\rm d}\cos(2\pi\phi_n)]x_n(1-x_n),
\label{eq:flx}\\
\phi_{n+1}
&=
\phi_n+\omega\pmod1,
\label{eq:flphi}
\end{align}
where $x_n\in[0,1]$ is the fiber coordinate, $\phi_n\in\T$ is the phase of the
quasiperiodic drive, $\alpha$ is the logistic control parameter,
$\epsilon_{\rm d}$ is the drive amplitude, and
\begin{equation}
\omega=\frac{\sqrt5-1}{2}.
\end{equation}
Following the standard parametrization of the forced-logistic SNA literature, we
use
\begin{equation}
\epsilon_{\rm d}'
=
\frac{\epsilon_{\rm d}}{4/\alpha-1}
=1,
\qquad
\epsilon_{\rm d}=\frac4\alpha-1.
\label{eq:epsd}
\end{equation}
Along this line a fractalization route from a smooth torus to an SNA occurs near
$\alpha_F\simeq2.6526$
\cite{PrasadMehraRamaswamy1998,PrasadNegiRamaswamy2001}.

The fiber Lyapunov exponent is
\begin{equation}
\Lambda_x
=
\lim_{N\to\infty}
\frac1N
\sum_{n=0}^{N-1}
\ln\left|
\alpha[1+\epsilon_{\rm d}\cos(2\pi\phi_n)](1-2x_n)
\right|.
\label{eq:LE}
\end{equation}
Representative numerical values are negative for the torus
($\alpha=2.50$, $\Lambda_x\simeq-0.216$), weakly negative for the SNA
($\alpha=2.70$, $\Lambda_x\simeq-0.025$), and positive for chaos
($\alpha=2.75$, $\Lambda_x\simeq+0.017$); only the sign and regime assignment are
used below.

For phase sensitivity define
\begin{equation}
q_n=\frac{\partial x_n}{\partial\phi_0},\qquad q_0=0.
\end{equation}
Differentiating Eq.~\eqref{eq:flx},
\begin{equation}
q_{n+1}=f_x(x_n,\phi_n)q_n+f_\phi(x_n,\phi_n),
\label{eq:q}
\end{equation}
with
\begin{align}
f_x(x,\phi)
&=
\alpha[1+\epsilon_{\rm d}\cos(2\pi\phi)](1-2x),\\
f_\phi(x,\phi)
&=
-2\pi\alpha\epsilon_{\rm d}\sin(2\pi\phi)x(1-x).
\end{align}
For one initial state,
\begin{equation}
\gamma_N(x_0,\phi_0)=\max_{0\le n\le N}|q_n|.
\end{equation}
The conventional phase-sensitivity function takes an attractor-wide minimum over
sampled initial states,
\begin{equation}
\Gamma_N=\min_{(x_0,\phi_0)\in\mathcal S}\gamma_N(x_0,\phi_0),
\label{eq:Gamma}
\end{equation}
where $\mathcal S$ denotes the sampled post-transient initial-state set.  Its
unbounded growth is an established SNA diagnostic
\cite{PikovskyFeudel1995,PrasadNegiRamaswamy2001}.

\subsection{Numerical sensing architecture}

The resource calculation isolates uncertainty in the initial forcing phase.  The
corresponding continuum uncertainty cell is
\begin{equation}
\mathcal U_{\delta}^{(\phi)}(x_0,\phi_0)
=
\{(x_0,\phi_0+s\!\!\pmod1): |s|\le\delta\},
\label{eq:phasecell}
\end{equation}
so the fiber coordinate is known exactly within this declared architecture and only the
initial phase is uncertain.  Equations~\eqref{eq:D}--\eqref{eq:DER} apply to this
cell just as they do to a metric ball.

Numerically, after a transient, a base state $(x_0^{(i)},\phi_0^{(i)})$ is selected
and the continuum cell is sampled at four signed offsets,
\begin{equation}
\phi_0^{(i,o)}
=
\phi_0^{(i)}+o\delta\pmod1,
\qquad
o\in\{-1,-1/2,1/2,1\}.
\label{eq:offset}
\end{equation}
Here $i$ labels the sampled base state and $\delta$ is the nominal phase
half-width.  The unperturbed trajectory has $o=0$.

The finite-ensemble nominal-trajectory deviation is
\begin{equation}
R_N^{\rm samp}(\delta)
=
\max_{i,o}\max_{0\le n\le N}
|x_n^{(i,o)}-x_n^{(i,0)}|.
\label{eq:Rsamp}
\end{equation}
After monotonizing the sampled relation in $\delta$, let
$\delta_{N,\rm samp}^*(\eta)$ be the largest sampled phase half-width for which
$R_N^{\rm samp}\le\eta$, where $\eta$ is the declared reference-trajectory
deviation tolerance.  The numerical resource is
\begin{equation}
B_N^{\rm num}(\eta)
=
\left[-\log_2\delta_{N,\rm samp}^*(\eta)\right]_+.
\label{eq:Bnum}
\end{equation}
The unit phase interval is the reference scale.  The notation $\eta$ is kept
distinct from the minimax tolerance $\eps$ because the numerical predictor is the
nominal trajectory rather than the cell-optimal minimax trajectory.  For the same
continuum phase-only cell, Eq.~\eqref{eq:DER} bounds the ideal minimax,
nominal-trajectory, and diameter errors within a factor of two in error scale.  The
finite set of base phases and four offsets used in $R_N^{\rm samp}$ introduce a
separate sampling approximation.

\section{Main result: three initial-precision resource classes}

\begin{figure}[htbp]
\centering
\includegraphics[width=0.31\linewidth]{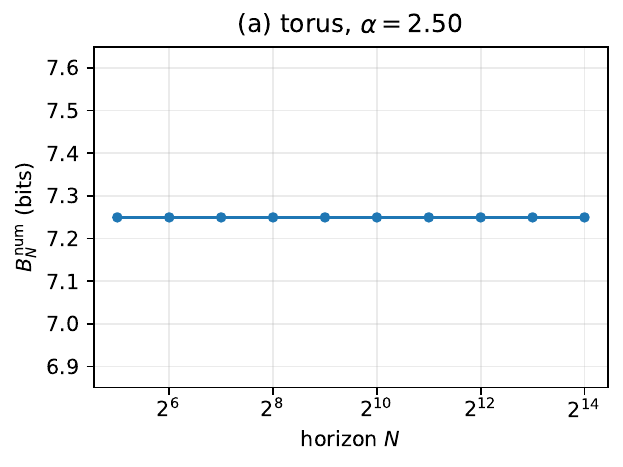}\hfill
\includegraphics[width=0.31\linewidth]{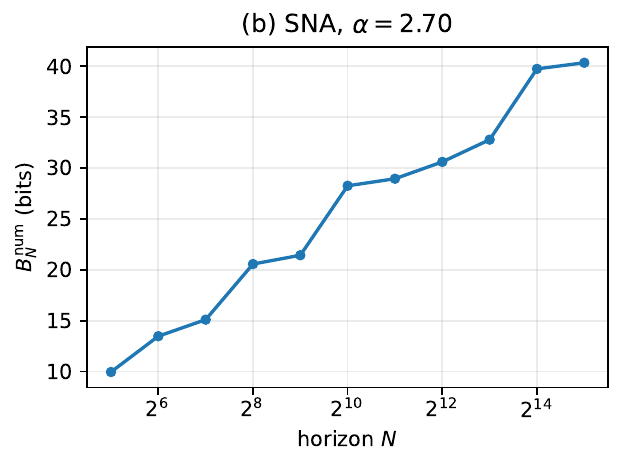}\hfill
\includegraphics[width=0.31\linewidth]{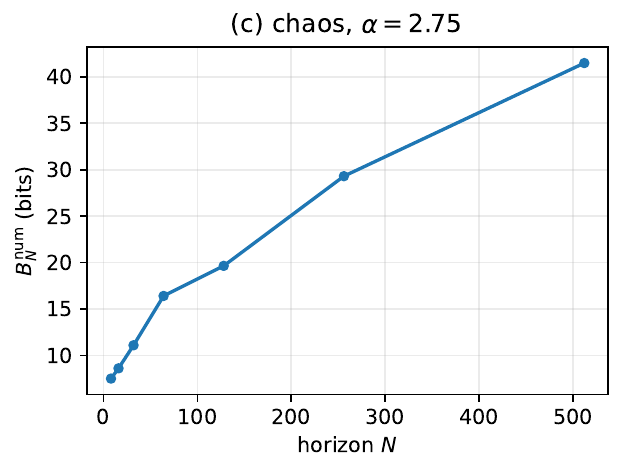}
\caption{Direct finite-perturbation precision cost at the same numerical
reference-trajectory tolerance $\eta=0.1$.  (a) Smooth torus:
$B_N^{\rm num}$ saturates.  (b) SNA: the horizontal axis is logarithmic and the
resource grows approximately linearly with $\log_2N$.  (c) Chaos: the horizontal
axis is linear and the resource grows approximately linearly with $N$ over the
range resolved before double-precision phase perturbations become limiting.
The panels use explicit marker/line encoding and remain interpretable in
monochrome.}
\label{fig:classes}
\end{figure}

At the common numerical tolerance $\eta=0.1$, Figure~\ref{fig:classes} compares
the three representative regimes on the axes that expose their observed horizon
dependence.  The torus saturates, the SNA is approximately linear when plotted
against $\log_2N$, and chaos is approximately linear in $N$ over the resolved
finite-perturbation window.  The figure therefore displays the central numerical
hierarchy directly: bounded, logarithmic-like, and linear-like precision demand.

Thus the numerical hierarchy is
\begin{equation}
\boxed{
\text{torus: bounded}
\quad\longrightarrow\quad
\text{SNA: logarithmic-like}
\quad\longrightarrow\quad
\text{chaos: linear-like}.
}
\label{eq:classes}
\end{equation}
The class labels in Eq.~\eqref{eq:classes} describe the resolved horizon windows
of the forced-logistic calculation and are not asymptotic classifications.  The
circle/Chebyshev benchmark in Eq.~\eqref{eq:chebB} is the analytic result; the three
forced-logistic scaling labels are finite-window numerical observations.

The SNA result is the central nontrivial case.  A negative fiber Lyapunov exponent
does not imply bounded initial phase precision.  Direct finite-perturbation fits
give logarithmic slopes of order a few bits per doubling of horizon, with the
coefficient depending on tolerance, fitting window, and worst-case aggregation.
The growth form is more robust than any one fitted coefficient.

\subsection{Why algebraic SNA sensitivity gives logarithmic precision demand}

The mechanism is simple once the operational resource is written explicitly.
Suppose that, over a local-response regime relevant to the prediction task, the
largest operational phase gain obeys
\begin{equation}
G_N^{\rm op}
\equiv
\sup_{(x_0,\phi_0)}
\max_{0\le n\le N}
\left|
\frac{\partial x_n}{\partial\phi_0}
\right|
\sim N^\mu,
\qquad \mu>0.
\label{eq:Gop}
\end{equation}
Then a small initial phase error $\delta$ produces
\begin{equation}
R_N(\delta)\sim G_N^{\rm op}\delta\sim N^\mu\delta.
\end{equation}
Keeping the phase-only nominal-trajectory error $R_N\lesssim\eta$ therefore
requires
\begin{equation}
\delta_{N,R}^*(\eta)\sim\eta N^{-\mu}.
\end{equation}
For the corresponding ideal reference-trajectory resource
$B_N^{(R)}(\eta)=[-\log_2\delta_{N,R}^*(\eta)]_+$, this gives
\begin{equation}
\boxed{
B_N^{(R)}(\eta)
\sim
\mu\log_2N+\log_2\frac1\eta+O(1).
}
\label{eq:sna_mech}
\end{equation}
Polynomial sensitivity becomes logarithmic precision cost because the resource is
the logarithm of the inverse admissible initial scale.

\begin{figure}[htbp]
\centering
\includegraphics[width=0.72\linewidth]{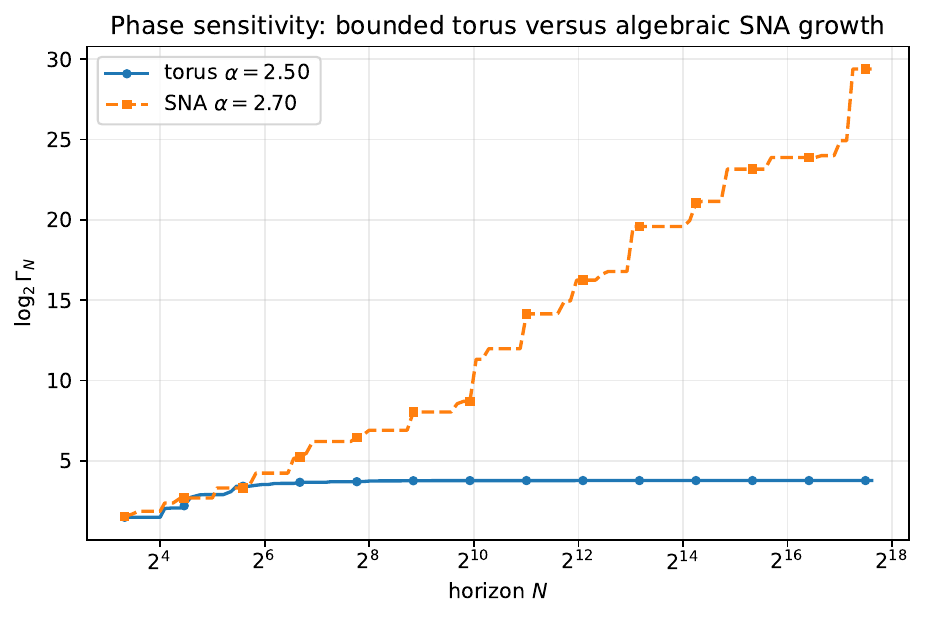}
\caption{Conventional phase-sensitivity function for the smooth torus and the
representative SNA.  The torus remains bounded, whereas the SNA displays
algebraic finite-window growth, which translates into logarithmic initial-precision
demand through Eq.~\eqref{eq:sna_mech}.}
\label{fig:phase}
\end{figure}

This relation explains the intermediate SNA class without identifying its
coefficient with a universal geometric dimension.  The conventional
Pikovsky--Feudel statistic $\Gamma_N$ in Eq.~\eqref{eq:Gamma} uses a minimum over
initial conditions, while $G_N^{\rm op}$ is worst-case in spirit.  Their exponents
and prefactors therefore need not coincide.  In the present data the conventional
SNA phase-sensitivity fit is algebraic over the resolved window
($\ln\Gamma_N\simeq\mu_{\rm PF}\ln N$ with
$\mu_{\rm PF}\simeq2.17$), whereas the direct resource gives a somewhat different
logarithmic coefficient.  The physically relevant statement is the shared
subexponential, algebraic character rather than equality of the two fitted
numbers.  Figure~\ref{fig:phase} shows this distinction directly: the torus
phase-sensitivity curve stays bounded over the resolved window, whereas the SNA
curve continues to grow algebraically, providing the sensitivity mechanism behind
Eq.~\eqref{eq:sna_mech}.

For ordinary chaos the corresponding local gain is exponential,
$G_N^{\rm op}\sim e^{\lambda N}$, and the same inversion gives
$B_N\sim(\lambda/\ln2)N+O(1)$, consistent with the exact Chebyshev calibration
and the direct chaotic numerics.

\section{Extreme-value convergence of the sampled worst case}

Because a finite maximum can underestimate a true supremum, especially on an SNA, we test the sampled worst case by nested phase-grid refinement.  For an $M$-point grid, let $R_{N,M}^{\max}(\delta)$ denote the maximum, over its base phases, of the horizon-wide deviation obtained from the four offsets in Eq.~\eqref{eq:offset}.  For each $\alpha$, $M_{\max}=4096$ equally spaced initial phases were propagated for $3\times10^4$ transient iterations from $x=0.2$ to
generate base states.  Nested dyadic subsets with
\begin{equation}
M=32,64,\ldots,4096
\end{equation}
were then evaluated using the four offsets in Eq.~\eqref{eq:offset}.
For the most sensitive $\delta=10^{-8}$ check the grid was extended to
$M=8192$.

\begin{figure}[htbp]
\centering
\includegraphics[width=0.72\linewidth]{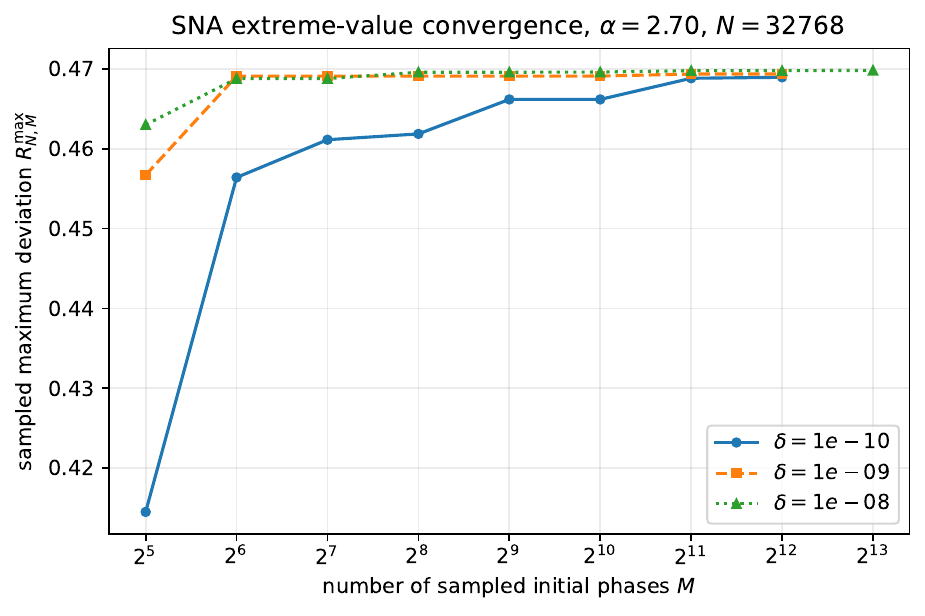}
\caption{Nested-grid convergence of the sampled SNA extreme value at
$\alpha=2.70$ and $N=32768$.  Different perturbation half-widths use distinct line
styles and markers.  The plateau of the sampled maximum with increasing phase
count substantially reduces, but does not mathematically eliminate, the possibility
that a narrower unresolved phase set produces a larger supremum.}
\label{fig:conv}
\end{figure}

For the SNA at $\alpha=2.70$ and $N=32768$, the sampled maximum at
$\delta=10^{-8}$ changes from $0.4630$ at $M=32$ to $0.46959$ at $M=256$,
$0.46979$ at $M=4096$, and $0.46982$ at $M=8192$.
The $4096\to8192$ change is about $7\times10^{-5}$ relative.
At $\delta=10^{-10}$, the larger change associated with rarer extreme events is
already only $2.5\times10^{-4}$ relative between $M=2048$ and $4096$.
For chaos at $\alpha=2.75$ and $N=4096$, the corresponding maxima stabilize near
$0.6163$, with changes below $4\times10^{-4}$ relative from $M=2048$ to $4096$
for all three tested perturbation scales.  Figure~\ref{fig:conv} visualizes the
SNA refinement directly: the sampled maximum rises at coarse phase resolution and
then forms a clear plateau, showing that the reported long-horizon envelope is
numerically stable to further phase-grid refinement.

This convergence study does not prove the mathematical supremum over phase.
It does, however, show that the long-horizon envelope near $0.47$ is not an artifact of a 16--256 point phase sample.  The complete
convergence table, including the chaotic case, is provided in Appendix~\ref{app:conv}.

\section{Long activation horizons near the torus--SNA onset}

The second main numerical result concerns the time needed for SNA sensitivity to
reach a finite operational scale.  Immediately above the reported fractalization
point, derivative-based phase sensitivity can already grow while a finite
perturbation remains microscopically small for extremely long horizons.

\begin{figure}[htbp]
\centering
\includegraphics[width=0.72\linewidth]{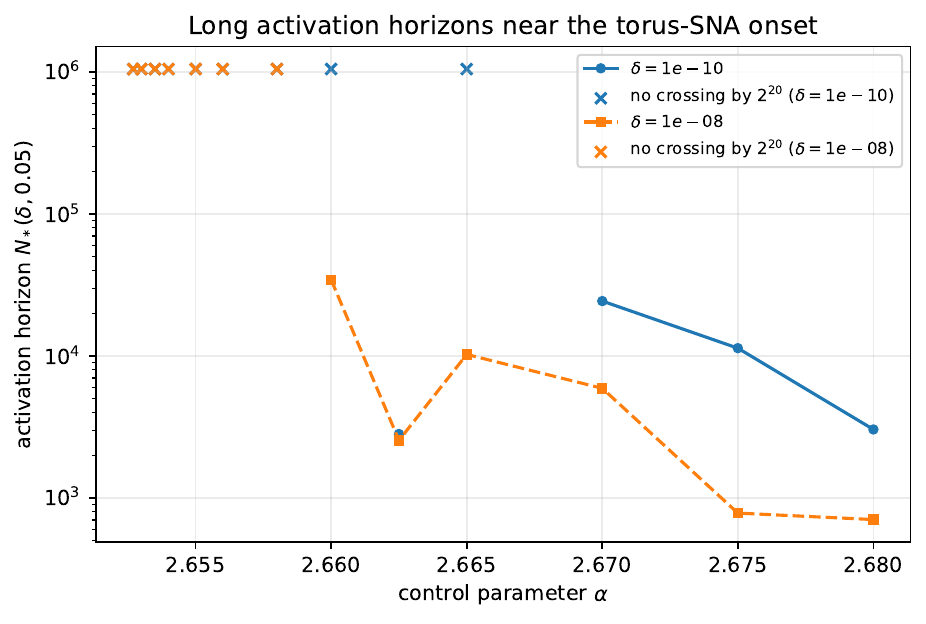}
\caption{First sampled horizon at which the reference-trajectory deviation reaches
$0.05$ near the torus--SNA onset.  Cross markers at $2^{20}$ denote runs with no
observed crossing by the maximum simulated horizon.  The large and nonmonotonic activation times separate derivative-based sensitivity
from finite-horizon operational visibility.}
\label{fig:activation}
\end{figure}

Define the sampled activation horizon
\begin{equation}
N_*(\delta,\eta)
=
\inf\{N:R_N^{\rm samp}(\delta)\ge\eta\},
\label{eq:Nstar}
\end{equation}
with $N_*=\infty$ if the threshold is not crossed during the simulated interval.
For $\delta=10^{-8}$ and $\eta=0.05$, several parameter values in
$2.6527\lesssim\alpha\lesssim2.658$ did not cross the operational scale by
\begin{equation}
N=1{,}048{,}576=2^{20}.
\end{equation}
At $\alpha=2.660$ the sampled crossing occurred near
$N\simeq3.4\times10^4$, while at $\alpha=2.6625$ it occurred near
$N\simeq2.6\times10^3$.  The dependence on $\alpha$ is strongly nonmonotonic, so
no critical power law is inferred from these data.  Figure~\ref{fig:activation}
makes both features visible: a broad near-onset interval remains unactivated up to
the maximum simulated horizon, while nearby parameter values can activate orders
of magnitude earlier.

The relevant classification is therefore not exhausted by a dynamical label.
For a finite-precision observer one must distinguish
\begin{equation}
\text{dynamical regime},\qquad
\text{required accuracy},\qquad
\text{available horizon}.
\end{equation}
A system can already be strange in an asymptotic diagnostic while remaining
operationally torus-like at a fixed finite tolerance for a very long time.

\section{Discussion}

The central result is the resource hierarchy in Eq.~\eqref{eq:classes}.
The exact circle/Chebyshev map shows that exponential instability produces a
linear initial-precision cost.  In the forced-logistic calculation, the representative
smooth torus shows bounded phase-precision demand over the resolved horizons, while
the representative SNA exhibits the intermediate logarithmic-like burden associated
with algebraic phase sensitivity.  This is an operational classification: it
does not redefine the underlying attractors, and it does not identify the fitted
resource coefficient with a universal fractal dimension or Lyapunov quantity.

Accuracy provides a second operational control, distinct from dynamical class.  Proposition~\ref{prop:scale} shows that fixed
accuracies inside one common scale-separable regime differ only by a constant bit
offset.  The fine-accuracy curves indeed show this approximate offset equivalence.
At much coarser tolerances, however, finite-scale structure can become
operationally irrelevant.  At fixed SNA dynamics ($\alpha=2.70$) the sampled
crossover is near $\eta\simeq0.47$, while at fixed chaotic dynamics
($\alpha=2.75$) it lies near $\eta\simeq0.615$--$0.620$ in the present
nominal-trajectory convention.  These are finite-horizon masking scales, not claimed asymptotic invariants; detailed scans are reported in Appendix~\ref{app:accuracy}.

The inverse horizon interpretation is useful experimentally.  If
\begin{equation}
B_N^{\rm num}(\eta)\simeq\mu_B\log_2N+C(\eta),
\end{equation}
with logarithmic slope $\mu_B>0$ and tolerance-dependent offset $C(\eta)$,
then in an SNA-like regime a fixed precision budget $B_0$ gives
\begin{equation}
N_{\max}\sim2^{[B_0-C(\eta)]/\mu_B}.
\end{equation}
One extra initial bit therefore multiplies the accessible horizon.  In a chaotic
regime with linear slope $\kappa_B>0$,
\begin{equation}
B_N^{\rm num}(\eta)\simeq\kappa_BN+C(\eta),
\end{equation}
one instead obtains
\begin{equation}
N_{\max}\sim\frac{B_0-C(\eta)}{\kappa_B},
\end{equation}
so additional bits extend the horizon additively.

The framework connects directly to several established ideas.  In particular, the phase-sensitivity diagnostic of SNAs \cite{PikovskyFeudel1995} is inverted here into a one-shot initial measurement requirement.  Finite-size Lyapunov exponents already
describe scale-dependent error growth \cite{Boffetta1998,Cencini2000}; here the
classified quantity is the initial resolution required by a fixed horizon-wide
task.  Predictive rate-distortion and metric mean dimension allow much more general
encodings \cite{MarzenCrutchfield2016,LindenstraussTsukamoto2018}, whereas the
present architecture intentionally keeps a local physical sensor.  Restoration
entropy concerns a continuing communication rate \cite{Kawan2021}; $B_N$ is paid
once, at the initial time.

Several limitations delimit the scope of these results.  The phase grid remains finite, so the true supremum is not proved.  The order of
$\delta\to0$ and $N\to\infty$ is delicate near the SNA onset.  The resource depends
on the declared sensing coordinate and metric; if the forcing phase is supplied by
an exact external clock rather than measured, phase precision is not the relevant
resource.  Finally, the present requirement is horizon-wide; terminal-time
prediction can behave differently in folded bounded maps.

\section{Conclusion}

Initial precision can be treated as an operational resource for prediction.
In a common scale-separable regime, changing a fixed tolerance changes only a
constant number of bits.  The exactly solvable circle/Chebyshev map provides the
calibration $B_N=N\log_2m+O(1)$ for ordinary exponential chaos.

The forced logistic map adds the nontrivial intermediate case.  Under a declared
phase-only sensing architecture, a smooth torus, an SNA, and chaos exhibit bounded,
logarithmic-like, and linear-like precision demand over the resolved horizons.
For the SNA, the mechanism is the conversion of algebraic phase sensitivity into
a logarithmic cost in inverse initial scale.  Nested phase grids up to thousands
of samples show that the extreme-value envelopes used in this conclusion are
numerically stable, while the near-onset calculations reveal activation horizons
that can exceed one million iterations.

Finite-precision prediction therefore depends on more than the asymptotic
dynamical label.  The measurement architecture, required accuracy, and horizon
determine which dynamical structure must actually be resolved.

\section*{Acknowledgments}

This work was supported by the research grant SP016 from SOBIN Institute LLC.

\appendix

\section{Accuracy masking at fixed dynamics}
\label{app:accuracy}

At fixed dynamics, coarse tolerances can mask a resource class that becomes visible
at finer tolerance.  These scans are secondary to the torus--SNA--chaos hierarchy
because the crossover values depend on finite horizon, sampling, and the
nominal-trajectory error convention.

\begin{figure}[htbp]
\centering
\includegraphics[width=0.72\linewidth]{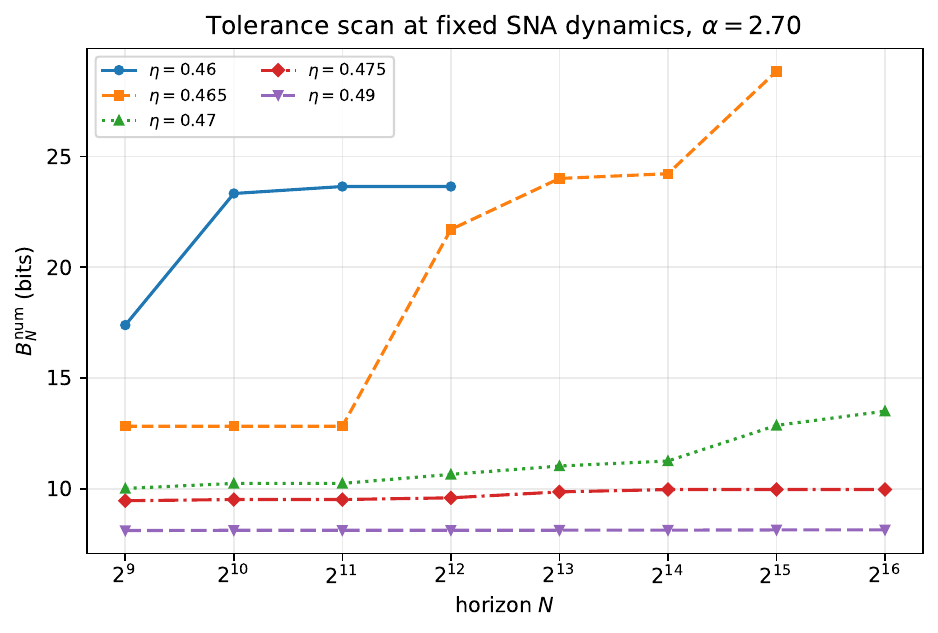}
\caption{Finite-horizon tolerance scan at fixed SNA dynamics.  Curves use distinct
line styles and markers.  The sampled crossover is near $\eta\simeq0.47$ in the
present reference-trajectory convention.}
\label{fig:snath}
\end{figure}

For the SNA at $\alpha=2.70$, the sampled crossover is near
$\eta\simeq0.47$.  Below it, the direct resource continues to grow over the
resolved horizon; sufficiently above it, the resource nearly saturates.
Figure~\ref{fig:snath} shows this masking transition: curves below the crossover
retain pronounced horizon dependence, whereas coarser tolerances flatten toward a
bounded-looking response over the same finite window.

\begin{figure}[htbp]
\centering
\includegraphics[width=0.72\linewidth]{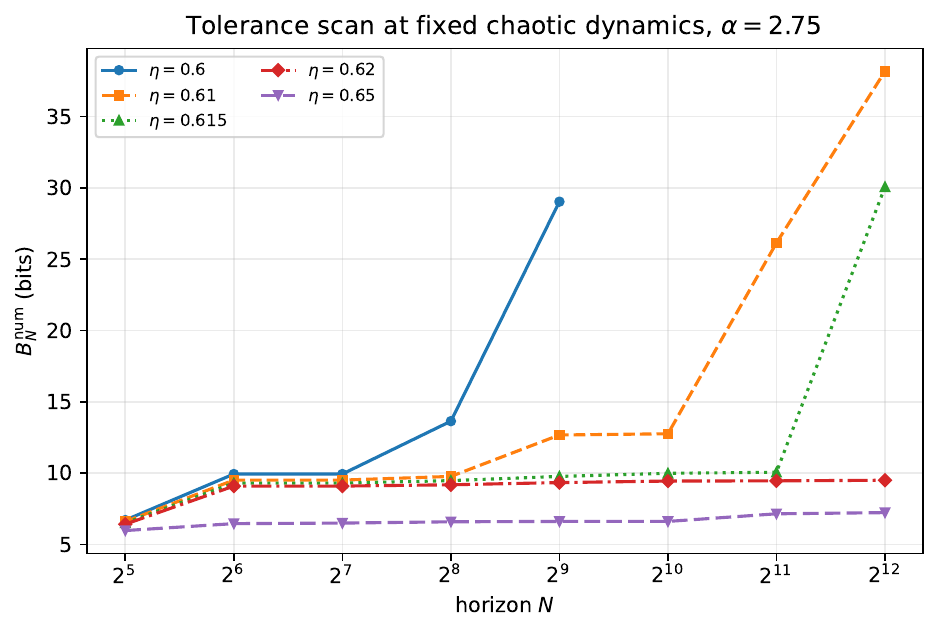}
\caption{Finite-horizon tolerance scan at fixed chaotic dynamics.  The sampled
crossover lies near $\eta\simeq0.615$--$0.620$.}
\label{fig:chth}
\end{figure}

For chaos at $\alpha=2.75$, the analogous sampled crossover lies near
$\eta\simeq0.615$--$0.620$.  Figure~\ref{fig:chth} shows the corresponding
finite-horizon masking in the chaotic regime: sufficiently coarse tolerances hide
the growth that is recovered once the tolerance resolves smaller-scale separation.

\section{Fine-accuracy offset equivalence}

\begin{figure}[htbp]
\centering
\includegraphics[width=0.72\linewidth]{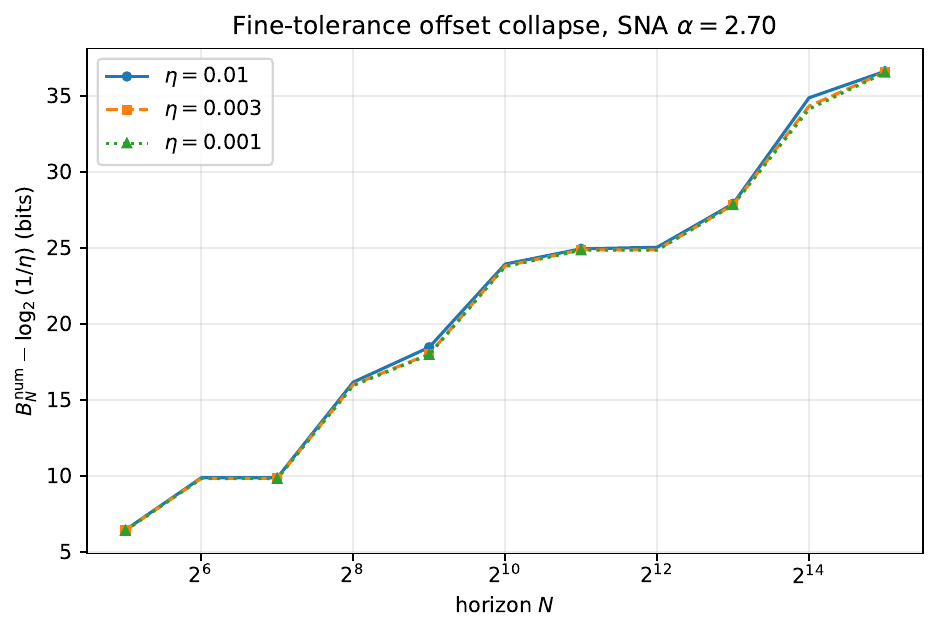}
\caption{SNA fine-accuracy offset collapse after subtracting
$\log_2(1/\eta)$.  The near-collapse supports a common local resource class for
these fine tolerances.}
\label{fig:offset}
\end{figure}

The same inversion used in Proposition~\ref{prop:scale} predicts a bounded
difference between two reference-trajectory resource curves when $R_N(\delta)$
remains in one scale-separable branch.  For
$\eta=10^{-2},3\times10^{-3},10^{-3}$, the measured mean bit difference between
the latter two tolerances is close to
\begin{equation}
\log_2 3=1.58496\ldots,
\end{equation}
with values approximately $1.586$ (torus), $1.596$ (SNA), and $1.589$ (chaos)
in the stored numerical test.  Figure~\ref{fig:offset} shows the SNA case after
subtracting the predicted tolerance offset: the three curves nearly collapse,
which is the expected signature of a common local scale-separable resource law.

\begin{figure}[htbp]
\centering
\includegraphics[width=0.72\linewidth]{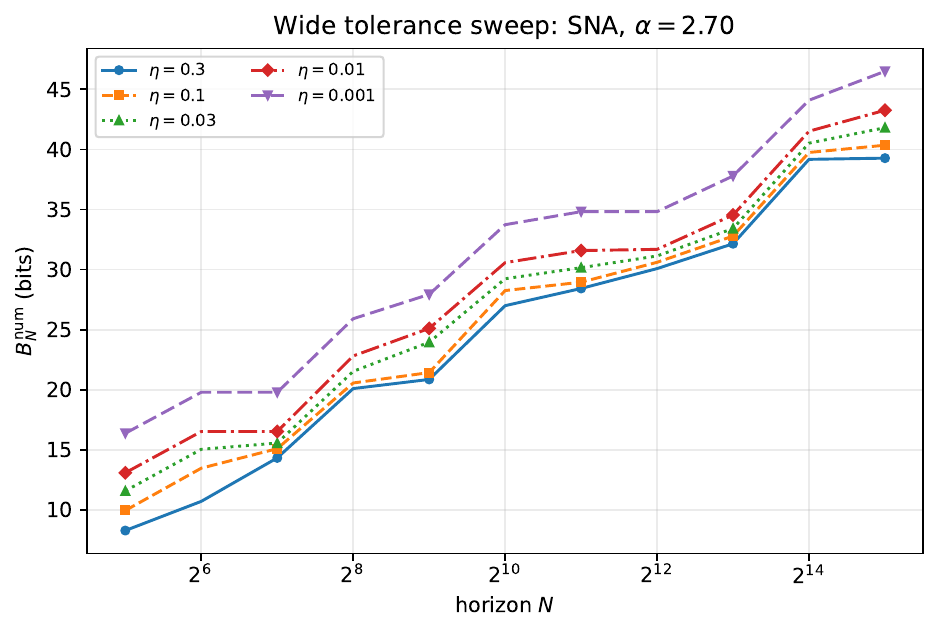}
\caption{Wide tolerance sweep at $\alpha=2.70$.}
\label{fig:wideSNA}
\end{figure}

\begin{figure}[htbp]
\centering
\includegraphics[width=0.72\linewidth]{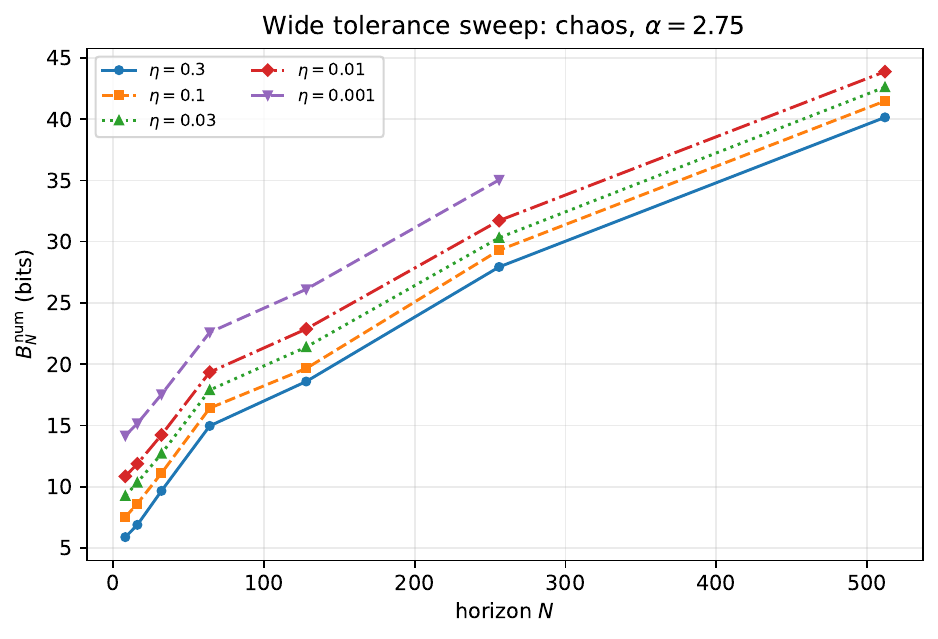}
\caption{Wide tolerance sweep at $\alpha=2.75$.}
\label{fig:wideChaos}
\end{figure}

The wider sweeps in Figs.~\ref{fig:wideSNA} and \ref{fig:wideChaos} place
that local offset equivalence in context.  In the SNA and chaotic cases alike,
fine tolerances preserve the underlying horizon-dependent burden, while sufficiently
coarse tolerances move into a masked finite-scale regime.

\section{Extreme-value convergence table}
\label{app:conv}

The nested-grid convergence calculation used equally spaced initial phases and
dyadic subsets, so every smaller grid is contained in the next larger grid.
Table~\ref{tab:conv} gives selected maximum deviations from the nested-grid calculation.

\begin{table}[htbp]
\centering
\caption{Selected nested-grid convergence values.}
\label{tab:conv}
\begin{tabular}{ccccc}
\toprule
$\alpha$ & $N$ & $\delta$ & $M$ & $R_{N,M}^{\max}$\\
\midrule
2.70 & 32768 & $10^{-10}$ & 256  & 0.46187\\
2.70 & 32768 & $10^{-10}$ & 2048 & 0.46884\\
2.70 & 32768 & $10^{-10}$ & 4096 & 0.46896\\
2.70 & 32768 & $10^{-8}$  & 256  & 0.46959\\
2.70 & 32768 & $10^{-8}$  & 4096 & 0.46979\\
2.70 & 32768 & $10^{-8}$  & 8192 & 0.46982\\
2.75 & 4096  & $10^{-10}$ & 256  & 0.61543\\
2.75 & 4096  & $10^{-10}$ & 4096 & 0.61628\\
2.75 & 4096  & $10^{-8}$  & 256  & 0.61611\\
2.75 & 4096  & $10^{-8}$  & 8192 & 0.61627\\
\bottomrule
\end{tabular}
\end{table}

\begin{figure}[htbp]
\centering
\includegraphics[width=0.72\linewidth]{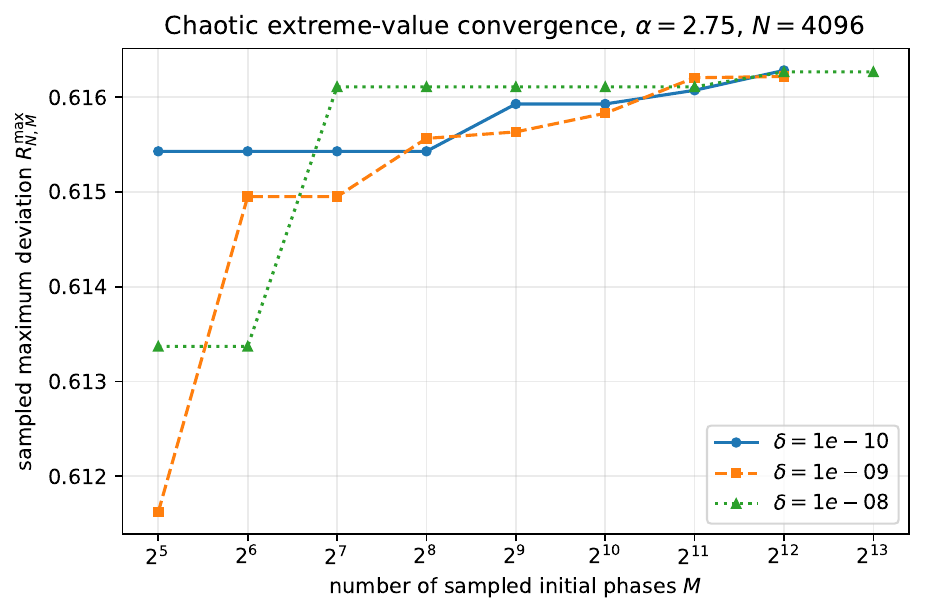}
\caption{Nested-grid extreme-value convergence for the chaotic representative.}
\label{fig:convchaos}
\end{figure}

Figure~\ref{fig:convchaos} gives the chaotic counterpart of the SNA convergence
test.  Its sampled maxima likewise approach a plateau with increasing phase count,
showing that the chaotic long-horizon envelope is not set by the coarsest phase
grid.

\section{Near-onset phase-sensitivity diagnostics}

\begin{figure}[htbp]
\centering
\includegraphics[width=0.72\linewidth]{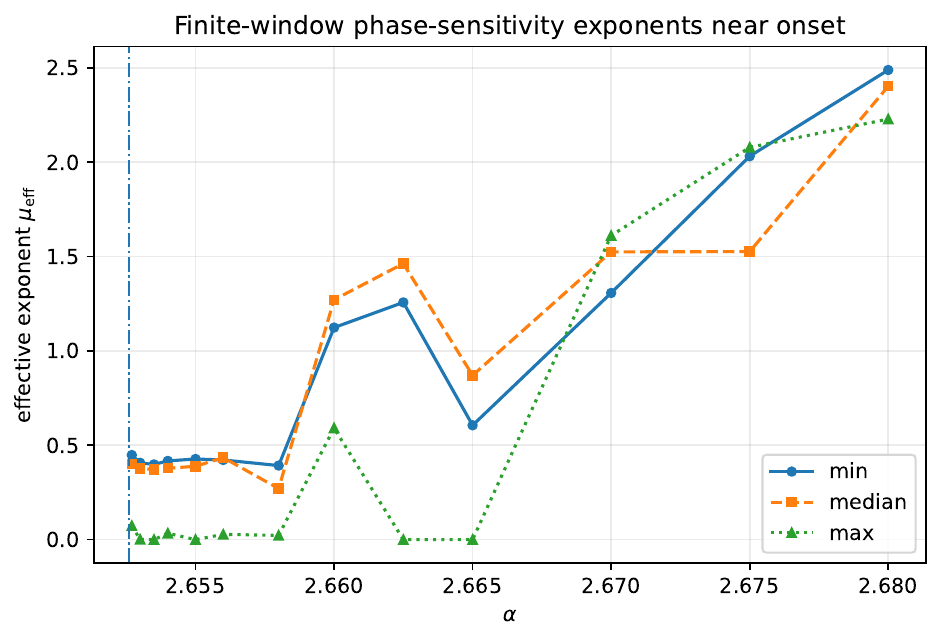}
\caption{Finite-window effective phase-sensitivity exponents near the torus--SNA
onset.  Distinct line styles and markers identify the aggregation statistics.}
\label{fig:onsetmu}
\end{figure}

Finite-window phase-sensitivity exponents vary strongly and nonmonotonically near
the fractalization threshold.  Minimum-, median-, and maximum-based summaries over
sampled initial phases need not agree because they answer different aggregation
questions.  Figure~\ref{fig:onsetmu} shows this separation explicitly: the three
aggregation rules can yield substantially different effective exponents at the same
control parameter, so no single near-onset exponent is treated as universal.

\section{Empirical parameter--tolerance map}

\begin{figure}[htbp]
\centering
\includegraphics[width=0.68\linewidth]{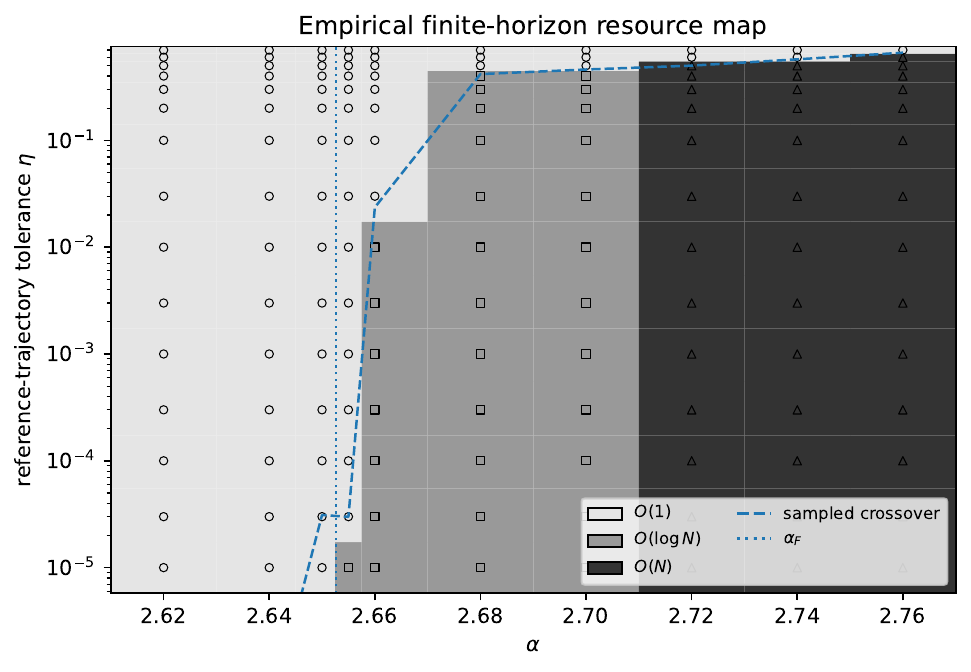}
\caption{Empirical finite-horizon resource map.  Grayscale levels and marker shapes
encode the three observed classes; the dashed curve is a sampled crossover estimate
and the dotted vertical line marks the literature fractalization threshold.}
\label{fig:phasemap}
\end{figure}

Figure~\ref{fig:phasemap} summarizes the finite-horizon numerical
classification over a wider $(\alpha,\eta)$ grid.  It should not be read as a
rigorous phase diagram: the boundary depends on the finite horizon and numerical
classification criterion.

\section{Numerical reproducibility notes}

The main resource scans used post-transient base states, four signed phase offsets
as in Eq.~\eqref{eq:offset}, and a monotone inversion of sampled maximum deviation
versus initial half-width.  The Lyapunov sweeps used long-time averages of
Eq.~\eqref{eq:LE}; phase-sensitivity recurrences used Eq.~\eqref{eq:q}.  Near-onset
activation scans followed selected perturbations to $2^{20}$ iterations.

The convergence calculation uses $3\times10^4$ transient iterations, equally
spaced phase grids, nested dyadic subsets, and the same four signed offsets as the
resource calculation.

\end{document}